\documentclass[letterpaper, 10pt, doublecolumn, conference]{ieeeconf}  

\IEEEoverridecommandlockouts 
\usepackage{tabu}
\usepackage{epsfig} 
\usepackage{amsmath, amsfonts, amssymb}
\let\labelindent\relax
\usepackage{enumitem}
\usepackage{pifont}
\usepackage[dvipsnames]{xcolor}
\usepackage{tikz}
\usetikzlibrary{arrows,shapes}
\usetikzlibrary{arrows.meta}
\usepackage[T1]{fontenc}
\usepackage{aecompl}

\usepackage{pgfplots}
\usepackage{algorithm2e}
\usepackage{varwidth}
\usepackage{soul}
\usepackage{verbatim}
\usepackage{cite}
\usepackage{booktabs}
\usepackage{float}

\usepackage{graphics} 
\usepackage{caption}
\usepackage{subcaption}
\usepackage{multirow}
\usepackage{setspace}
\usepackage{graphicx}

\usepackage{amsthm}
\usepackage[scientific-notation=true]{siunitx}
\usepackage[colorlinks]{hyperref}
\usepackage{amsfonts}
\usepackage{dsfont}
\usepackage{placeins}
\definecolor{darkblue}{RGB}{0,101,204}
\definecolor{carorange}{RGB}{255,131,0}
\newtheorem{theorem}{Theorem}

\theoremstyle{definition}
\newtheorem{definition}{Definition}
\newtheorem{assumption}{Assumption}
\theoremstyle{remark}
\newtheorem{remark}{Remark}

\newcounter{tmp}

\title{\LARGE \bf
To Stay or to Bypass: Unraveling Mainline Vehicles' Aggregate Strategic Decision-Making at Highway Weaving Ramps}

\author{Haohui He, Kexin Wang, and Ruolin Li$^{1}$
\thanks{$^{1}${H. He, K. Wang and R. Li are with Sonny Astani
Department of Civil and Environmental Engineering, University of Southern California, CA, USA.
{\tt\small haohuihe@usc.edu},
{\tt\small kwang255@usc.edu}
{\tt\small ruolinl@usc.edu}.
}
}
}

\begin{document}
\maketitle

\thispagestyle{empty}
\pagestyle{empty}


\begin{abstract}


The weaving ramp scenario is a critical bottleneck in highway networks due to conflicting flows and complex interactions among merging, exiting, and through vehicles. In this work, we propose a game-theoretic model to capture and predict the aggregate lane choice behavior of mainline through vehicles as they approach the weaving zone. Faced with potential conflicts from merging and exiting vehicles, mainline vehicles can either bypass the conflict zone by changing to an adjacent lane or stay steadfast in their current lane. Our model effectively captures these strategic choices using a small set of parameters, requiring only limited traffic measurements for calibration. The model's validity is demonstrated through SUMO simulations, achieving high predictive accuracy. The simplicity and flexibility of the proposed framework make it a practical tool for analyzing bottleneck weaving scenarios and informing traffic management strategies.

\end{abstract}
\section{Introduction}\label{intro}

\begin{figure*}
\centering
\includegraphics[width = \textwidth]{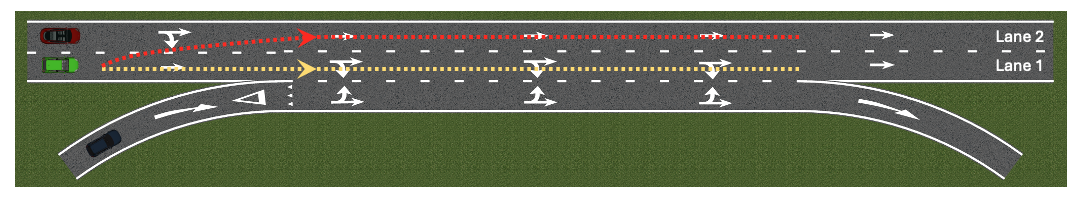}
\caption{In the vicinity of a highway weaving ramp, mainline through vehicles (indicated by green vehicles) must decide between two strategies: remaining in their current lane (\textit{steadfast} behavior, indicated by yellow traces) or shifting to an adjacent lane to bypass the potential conflicts (\textit{bypassing} behavior, indicated by red traces). The weaving zone involves vehicles merging onto the mainline from the on-ramp and vehicles diverging to leave the highway, creating potential conflicts that influence lane choice decisions. Our model captures the strategic behavior of mainline through vehicles by accurately predicting the proportion of steadfast and bypassing vehicles while maintaining reasonable computational complexity and
minimal calibration requirements.}
\label{fig:weaving_ramp}
\end{figure*}

In modern transportation systems, improving traffic efficiency has become a critical priority, with traffic congestion posing a growing challenge. According to \cite{inrix2024scorecard}, in 2023, the average U.S. driver spent 42 hours in congestion, approximately equivalent to a full work week, resulting in an estimated individual time loss for the driver valued at \$733 and a total economic cost exceeding \$70.4 billion nationwide. 
  
Frequent vehicle interactions are a significant contributor to traffic inefficiency, complicating drivers’ decision-making processes and disrupting the smooth traffic flow. A prominent interaction example is the weaving ramp scenario, where the close proximity of on-ramps and off-ramps creates conflicting movements between vehicles merging onto the highway and those preparing to exit. These overlapping paths lead to complex interactions, often resulting in congestion or even gridlock, which can severely degrade traffic conditions in the affected lanes. Studies such as ~\cite{daganzo2002behavioral,laval2006lane,chen2018capacity,rios2016automated} have investigated such inefficiency and its underlying cause from an analytical perspective, while~\cite{munoz2002bottleneck, srivastava2013empirical, lee2008empirical} focused on empirical investigations to discuss this topic. 

Traditionally, lane-changing behavior has been studied from a microscopic perspective, focusing on the detailed interactions between individual vehicles. 
Many studies such as~\cite{gipps1986model, ahmed1999modeling, hidas2002modelling, shi2022integrated}, focused on the influence of factors such as speed, acceleration, and other vehicle mechanical characteristics in relation to surrounding traffic. Other studies~\cite{jula2000collision, yang2018dltp} emphasize physical constraints, such as longitudinal spacing or voids between vehicles. Despite their effectiveness in capturing individual vehicle behaviors and generating high-fidelity trajectories, microscopic models are often limited in scalability, particularly in long-term, system-level predictions, and typically require extensive calibration with detailed data. 

In contrast, macroscopic models capture the aggregate dynamics of traffic flow considering the collective behavior of vehicle populations, making them suitable for system-level analysis. Ouyang et al.~\cite{ouyang2023lanechange} evaluated the effects of traffic flow conditions on lane change rates on weaving ramps, providing macroscopic guidelines to improve traffic efficiency and safety. Jin~\cite{jin2010kinematic} proposed a model that introduces a lane-changing intensity variable to represent the lateral impacts of lane changes on overall traffic flow, and further extended this theory in~\cite{jin2010macroscopic} to analyze lane-changing effects at the system level without relying on individual vehicle behaviors. Zhu et al.~\cite{zhu2022multilane} developed a flow-level coordination strategy that integrates lane-changing rules to improve ramp merging efficiency and prevent congestion. Furthermore, Arman et al.~\cite{arman2024macroscopic} proposed a macroscopic lane-changing model that requires only a small fraction (1-2\%) of vehicle trajectory data and eliminates the need for information from surrounding vehicles, thereby achieving a highly consistent lane balance. Overall, macroscopic models demonstrate clear advantages in enhancing system efficiency while simplifying modeling requirements through limited data and avoiding individual behavior analysis.

Beyond traditional modeling approaches, the game-theoretic method is an effective way to frame lane-changing decisions as strategic interactions between vehicles. Wang et al.~\cite{wang2015game} employed a differential game to jointly design the cost functions and acceleration controls for lane-changing maneuvers, based on anticipated behaviors of surrounding vehicles. Xu et al.~\cite{xu2024egtml} combined evolutionary game theory and machine learning to model mandatory lane-changing behaviors, capturing the progressive, adaptive interactions among drivers. Fu et al.~\cite{fu2025cooperative} designed a three-player, transferable-utility game framework that optimizes lane-changing strategies in a way that benefits all participants in most scenarios. While game theory provides a robust framework for representing the strategic aspects of lane choice, applying it at an aggregate level remains challenging due to the diverse decision-making processes involved, leading to a trade-off between model accuracy and computational efficiency. Nevertheless, modeling lane-choice behavior through game-theoretic approaches from a macroscopic perspective, enables us to consider vehicle interactions and overall traffic efficiency simultaneously, which can provide a novel perspective for traffic analysis.

This paper presents a novel macroscopic game-theoretic model, extending our previous work by explicitly incorporating increased scenario complexity. In our earlier studies~\cite{ruolin2019bifurcating, ruolin2020onramp}, we developed the Wardrop-like~\cite{wardrop1952some} frameworks, where vehicles select behaviors that minimize their individual costs, to model aggregate lane choice behavior across different traffic scenarios, including highway diverges with bifurcating lanes and at the vicinity of on-ramps. The framework also shows effectiveness in vehicles’ last-moment bypassing behavior at diverges~\cite{mehr2021diverge}. However, weaving ramps present more complex interactions for through vehicles due to merging and exiting flows, which has not been explicitly modeled and analyzed. Therefore, we extend this framework to handle the unique challenges posed by weaving ramp scenarios in following sections.

An illustration of the scenario is provided in Fig.~\ref{fig:weaving_ramp}. Through vehicles traveling in the lane adjacent to the ramps are significantly affected by interactions with both merging vehicles from the on-ramp and exiting vehicles heading toward the off-ramp. These vehicles must traverse the interaction zone to reach their destinations, creating high lane utilization and introducing potential conflicts. As a result, through vehicles are faced with two strategic choices: \textbf{either to continue traveling in their current lane (remaining \textit{steadfast}) or to shift to an adjacent lane farther from the ramps (\textit{bypassing}) to avoid potential congestion and delays.}

Given the previous discussed geometry features, modeling lane choice behavior in the weaving ramp scenario becomes significantly more complex. This complexity primarily stems from the increased number and diversity of vehicle interactions, which complicates the development of accurate cost models. Additionally, the increased complexity demands a more extensive and representative dataset to capture diverse interactions for reliable calibration and validation. These challenges are addressed in the subsequent sections, and the key contributions of this paper are summarized as follows: (1) We propose \textbf{a macroscopic game-theoretic model} for predicting the lane choice behavior of through vehicles in highway weaving ramp scenarios. (2) We validate \textbf{the accuracy of our model} using data obtained from extensive simulations under a range of weaving ramp conditions. (3) We validate \textbf{the robustness of our model} by calculating the mean prediction error rate of the cost function calibrated by the original data points in
different experimental scenarios. (4) We investigate the \textbf{practical implications} of each model parameter by recalibrating the model under various experimental conditions.

This paper is organized as follows. In Section~\ref{sec:model}, we propose the mathematical model of lane choice for through vehicles. In Section~\ref{sec:eq_prop}, we highlight the nice properties of the model by proving the existence and uniqueness of the calculated equilibrium. In Section~\ref{sec:simulation studies}, we use SUMO to generate data for the calibration, validation and analysis of our lane choice model. Finally, in Section~\ref{sec:future}, we summarize and draw the conclusions of this study and introduce our future work.



\section{The Strategic Lane Choice Model}\label{sec:model}
In this methodology section, we first formally define the research question. Then, we present the detailed formulation of the cost models and lane choice models used to capture through vehicles' strategic decision-making.

\subsection{The Lane Choice Behavior of Interest}

The weaving ramp section under study consists of two lanes in the mainline, Lane 1 and Lane 2, as well as an auxiliary lane, Lane 0, connecting an on-ramp and an off-ramp, as illustrated in Fig.~\ref{fig:weaving_ramp}.
Entering vehicles travel along the on-ramp and Lane 0, aiming to merge into Lane 1 within a constrained distance. Additionally, exiting vehicles typically shift from Lane 2 into Lane 1 and eventually shift into Lane 0 to take the off-ramp and exit the highway. 

Lane 1 is the outermost mainline lane adjacent to Lane 0 and serves as the immediate conflict zone for interactions among entering, exiting and mainline vehicles. Lane 2 is another mainline lane, providing an alternative route for Lane 1 vehicles to bypass the conflict zone. \textbf{Mainline vehicles in Lane 1 face a strategic decision in response to the complex interaction in the conflict zone. They can either stay steadfast in Lane 1, accepting potential interactions and possible disruptions from both entering and exiting vehicles, or shift to Lane 2 to bypass the conflict zone entirely. Through vehicles on Lane 1 evaluate the costs under both options and choose the one that minimizes the cost.} 
Vehicles initially positioned on Lane 0 or Lane 2 determine their lane-changing behavior based on their predefined travel intentions, such as merging onto or exiting from the mainline. The proportion of Lane 1 vehicles choosing to either remain steadfast or bypass to Lane 2 significantly shapes the overall traffic dynamics in the weaving zone. A large proportion of steadfast vehicles can create bottlenecks, leading to increased merging difficulty and risky maneuvers by entering vehicles, as well as complicating the lane-changing maneuvers of exiting vehicles from Lane 2. Conversely, if more Lane 1 vehicles bypass to Lane 2, the mainline flow near the ramps becomes smoother, but the congestion may shift to Lane 2, affecting through vehicles and creating new conflicts for those preparing to exit. Thus, understanding and predicting such lane choice behavior is critical for designing effective strategies to improve traffic conditions near weaving ramps.

Before delving into the detailed mathematical model, we first introduce some relevant notations to characterize the traffic flows in the weaving zone. The following lists different vehicle types and their flow rates: (1) Entering vehicles: $f_{\text{enter}}$, representing the flow of vehicles merging from the on-ramp into Lane 1; (2) Exiting vehicles: $f_{\text{exit}}$, representing the flow of vehicles that intend to exit the highway, typically shifting from Lane 2 into Lane 1, and eventually into the off-ramp; (3) Through vehicles on Lane 2: $f_2$, representing the flow of through vehicles traveling on Lane 2, unaffected by the merging or exiting interactions in the weaving zone; (4) Through vehicles on Lane 1: $f_1$, representing the total flow of through vehicles traveling on Lane 1 before the weaving zone, which will be split into those who choose to stay steadfast or bypass the conflict zone.

We further define the \textit{normalized flow ratios} as follows:
\begin{align}
    n_{\text{enter}} &:= \frac{f_{\text{enter}}}{ f_{\text{enter}} + f_{\text{exit}} + f_2}, \\
    n_{\text{exit}} &:= \frac{f_{\text{exit}}}{ f_{\text{enter}} + f_{\text{exit}} + f_2},\\
    n_2 &:= \frac{f_2}{ f_{\text{enter}} + f_{\text{exit}} + f_2},
\end{align}
where naturally, $n_{\text{enter}} + n_{\text{exit}} + n_2 = 1.$ The normalized flow ratios $n_{\text{enter}}, n_{\text{exit}}$ and $n_2$  indicate the proportion of entering, exiting, and through vehicles relative to the combined flow in the vicinity of Lane 1. The normalized flow ratios are considered known in the model, and are collected in the flow configuration vector $\mathbf{n} := \left(n_{\text{enter}}, n_{\text{exit}}, n_2\right).$

Next, we focus on the lane choice behavior of through vehicles on Lane 1. Let $f_1^{\text{s}}$ denote the flow of Lane 1 vehicles that stay \textit{steadfast} in Lane 1, preparing to interact with merging and exiting vehicles. Let $f_1^{\text{b}}$ denote the flow of Lane 1 vehicles that \textit{bypass} the weaving area by shifting to Lane 2. The proportion of each type of behavior on Lane 1 can be defined as:
\begin{align}
    x_1^\text{s} := \frac{f_1^\text{s}}{f_1}, \\
    x_1^\text{b} := \frac{f_1^\text{b}}{f_1},
\end{align}
where \( x_1^\text{s} \) represents the proportion of steadfast vehicles on Lane 1, and \( x_1^\text{b} \) represents the proportion of bypassing vehicles on Lane 1. A flow distribution vector $\mathbf{x}:=\left(x_1^\text{s}, x_1^\text{b}\right)$ is derived using our model and any feasible candidate must satisfy the following constraints:
\begin{align}\label{eq:flowcons}
    &x_1^\text{s} + x_1^\text{b} = 1, \\
    &x_1^\text{s} \geq 0,\ x_1^\text{b} \geq 0.
\end{align} 

In the subsequent sections, we introduce the cost models and the aggregate lane choice model, which determines the equilibrium distribution of lane choices of Lane 1 vehicles in the weaving zone. Mathematically, this framework predicts the flow distribution vector \( \mathbf{x} \) for a given flow configuration vector \( \mathbf{n} \), capturing the cost-based decision-making process of Lane 1 vehicles.

\subsection{The Cost Model}

To model the costs associated with the two lane choice options ($steadfast$ and $bypass$) of through vehicles on Lane 1, some key conditions must be satisfied. First, from an aggregate perspective, through vehicles choosing the same option should have the same cost. Second, the cost should be an increasing function of the relevant traffic flows, reflecting the principle that higher traffic volumes lead to greater travel difficulty and, consequently, higher costs. More specifically, a cost function can consist of two components: a traversing cost, which is proportional to the total flow on the chosen lane, and a merging cost, which is proportional to the product of the flows of merging parties.

The specific structure and parameter values of the cost model are determined through calibration and validation using data collected from simulations or real-world traffic datasets. The model is considered appropriate only when the calibration and validation results demonstrate satisfactory accuracy. For the sake of brevity, we omit the iterative design process and present only the final validated models in the following sections.

Let \( J_1^\text{s} \) denote the cost for through vehicles choosing the steadfast option, and \( J_1^\text{b} \) denote the cost for through vehicles choosing the bypassing option. We have
\begin{align}\label{eq:Js}
J_1^\text{s}(\mathbf{x}) &= C_1^\text{t} \left( \alpha x_1^\text{s}+\beta n_\text{exit}+ n_\text{enter}\right) \nonumber\\
&\ \ \ \ \ \ \ \ \ \ \ \ \ \ \ \ \ \ \ \ +C_1^\text{m} \left(\omega x_1^\text{s} n_\text{exit}+ x_1^\text{s} n_\text{enter}\right),\\
J_1^\text{b}(\mathbf{x}) &= C_2^\text{t}\left(\gamma x_1^\text{b}+ n_2\right) + C_2^\text{m} \left( \rho x_1^\text{b} n_2 + \delta x_1^\text{b} n_\text{exit} \right).
\end{align}

In the cost model, let \( C_i^\text{t} \) denote the unit traversing cost for Lane \( i \), where \( i \in \{1, 2\} \). After the strategic decisions of through vehicles, Lane 1 is shared by steadfast vehicles, exiting vehicles, and entering vehicles, while Lane 2 is occupied by bypassing vehicles and vehicles always traveling along Lane 2. The weight parameters \( \alpha, \beta, \gamma \) are assumed to be positive, representing the relative impact on the cost compared to a standard traversing behavior. If a parameter value exceeds 1, it indicates a higher cost relative to normal traversing, possibly due to lower speeds or increased discomfort. Conversely, values less than 1 suggest a lower impact, while a value of 1 indicates a neutral, baseline cost. Similarly, let \( C_i^\text{m} \) represent the unit merging cost for Lane \( i \), with \( \omega, \rho, \delta \) denoting the additional effort or discomfort associated with conflicting merging movements. A higher value of \( \delta \) implies a more challenging merging process, reflecting the increased difficulty compared to standard merging behavior. Then let the cost coefficient vector be defined as \(\mathbf{C} := (C_i^\text{t}, C_i^\text{m}, \alpha, \beta, \omega, \gamma,\rho, \delta \, | \, i \in \{1, 2\})\), which includes all the parameters that need be calibrated prior to the deployment on a new weaving ramp zone.

\begin{remark}[Application Scenarios of Our Model]
    The proposed cost functions, which are proportional to flow rates, are particularly well-suited for modeling uncongested scenarios where traffic flows smoothly. However, they may not fully capture the dynamics of congested conditions, such as queue formation, and thus are less applicable in oversaturated environments.
\end{remark}

\subsection{The Strategic Lane Choice Model}

We are now ready to present the formal decision-making model for through vehicles' lane choice in the weaving ramp scenario, inspired by the classical Wardrop Equilibrium Conditions~\cite{wardrop1952some}.

\begin{definition}[Strategic Lane Choice Equilibrium]\label{def:wdp_basic}
For a given weaving ramp configuration \( G = (\mathbf{N}, \mathbf{C}) \), a flow distribution vector \( \mathbf{x} \) is in equilibrium if and only if

\begin{equation}\label{eq:eq_def}
    \begin{aligned}
        x_1^\text{s}  (J_1^\text{s}(\mathbf{x}) - J_1^\text{b}(\mathbf{x})) &\leq 0 ,\\
        x_1^\text{b}  (J_1^\text{b}(\mathbf{x}) - J_1^\text{s}(\mathbf{x})) &\leq 0.
    \end{aligned}
\end{equation}
\end{definition}

This equilibrium model encodes the conditions under which each vehicle selects the option that minimizes its own cost. Specifically, if bypassing is less costly, i.e., \( J_1^\text{s}(\mathbf{x}) - J_1^\text{b}(\mathbf{x}) > 0 \), all Lane 1 vehicles will choose to bypass, resulting in \( x_1^\text{s} = 0 \). Conversely, if remaining steadfast is less costly, i.e., \( J_1^\text{b}(\mathbf{x}) - J_1^\text{s}(\mathbf{x}) > 0 \), all Lane 1 vehicles will stay in their current lane, leading to \( x_1^\text{b} = 0 \). If the costs of both strategies are equal, i.e., \( J_1^\text{b}(\mathbf{x}) - J_1^\text{s}(\mathbf{x}) = 0 \), then a mixture of both behaviors can occur, resulting in non-zero values for \( x_1^\text{s} \) and \( x_1^\text{b} \).

These conditions can also be interpreted in reverse based on observed lane choices: If no vehicles choose to bypass (\( x_1^\text{b} = 0 \)), it implies that bypassing is more costly than remaining steadfast (\( J_1^\text{b}(\mathbf{x}) - J_1^\text{s}(\mathbf{x}) > 0 \)). If all vehicles bypass (\( x_1^\text{s} = 0 \)), it suggests that remaining steadfast is more costly (\( J_1^\text{s}(\mathbf{x}) - J_1^\text{b}(\mathbf{x}) > 0 \)). If both \( x_1^\text{s} \) and \( x_1^\text{b} \) are non-zero, the costs must be equal, i.e., \( J_1^\text{b}(\mathbf{x}) - J_1^\text{s}(\mathbf{x}) = 0 \).

\section{Existence and Uniqueness of the Equilibrium}\label{sec:eq_prop}

The aggregate lane choice equilibrium implies that each vehicle selects the option that minimizes its cost, resulting in one of three outcomes: (1) all vehicles remain steadfast, (2) all vehicles bypass, or (3) a mixture of both behaviors if the costs are equal. We now establish the existence and uniqueness of this equilibrium.

\begin{theorem}[Existence and Uniqueness]\label{thm:uniq}
For any given weaving ramp configuration \( G = (\mathbf{N}, \mathbf{C}) \), the equilibrium flow distribution vector \( \mathbf{x} \) defined in Definition~\ref{def:wdp_basic} always exists and is unique.
\end{theorem}

\begin{proof}
To prove this, we first write the cost functions for through vehicles as follows:
\begin{align}
J_1^\text{s}(\mathbf{x}) &= C_1^\text{t} \left( \alpha (1 - x_1^\text{b}) + \beta n_\text{exit} + n_\text{enter} \right) \nonumber \\
&\ \ \ \  + C_1^\text{m} \left(\omega  (1 - x_1^\text{b}) n_\text{exit} + (1 - x_1^\text{b}) n_\text{enter} \right), \\
J_1^\text{b}(\mathbf{x}) &= C_2^\text{t} \left( \gamma x_1^\text{b} + n_2 \right) + C_2^\text{m} \left( \rho x_1^\text{b} n_2 + \delta x_1^\text{b} n_\text{exit} \right). \nonumber
\end{align}

\noindent Next, we analyze the behavior of the cost functions with respect to \( x_1^\text{b} \):

\begin{itemize}
    \item \( \frac{\partial J_1^\text{s}}{\partial x_1^\text{b}} = - C_1^\text{t} \alpha - C_1^\text{m} (\omega n_\text{exit} + n_\text{enter}) \), which is a negative constant, making \( J_1^\text{s}(\mathbf{x}) \) a decreasing affine function of \( x_1^\text{b} \).
    \item \( \frac{\partial J_1^\text{b}}{\partial x_1^\text{b}} = C_2^\text{t} \gamma + C_2^\text{m} (\rho n_2 + \delta n_\text{exit}) \), which is a positive constant, making \( J_1^\text{b}(\mathbf{x}) \) an increasing affine function of \( x_1^\text{b} \).
\end{itemize}
The above properties ensure that the cost functions can intersect at most once, which directly implies that the equilibrium is unique. Let us consider three scenarios based on the relative magnitudes of \( J_1^\text{s}(\mathbf{x}) \) and \( J_1^\text{b}(\mathbf{x}) \):

\begin{itemize}
    \item Case (a): For all \( x_1^\text{b} \in [0, 1] \), \( J_1^\text{s}(\mathbf{x}) > J_1^\text{b}(\mathbf{x}) \). This implies that all vehicles will choose to bypass, resulting in \( x_1^\text{s} = 0 \) and \( x_1^\text{b} = 1 \).
    
    \item Case (b): \( J_1^\text{s}(\mathbf{x}) \) and \( J_1^\text{b}(\mathbf{x}) \) intersect at a unique point \( \bar{x}_1^\text{b} \in (0, 1) \). This intersection defines the unique equilibrium distribution, where \( x_1^\text{b} = \bar{x}_1^\text{b} \) and \( x_1^\text{s} = 1 - \bar{x}_1^\text{b} \).
    
    \item Case (c): For all \( x_1^\text{b} \in [0, 1] \), \( J_1^\text{s}(\mathbf{x}) < J_1^\text{b}(\mathbf{x}) \). In this scenario, all vehicles will remain steadfast, resulting in \( x_1^\text{s} = 1 \) and \( x_1^\text{b} = 0 \).
\end{itemize}

Thus, in all cases, the equilibrium exists and is unique. This concludes the proof.
\end{proof}

In the following sections, we validate the model using simulation data to illustrate its predictive capabilities and performance in various weaving ramp configurations.

\section{Model Validation} \label{sec:simulation studies}

To validate our model, we use data collected from the microscopic traffic simulator SUMO~\cite{SUMO2012}. An overview of the simulated highway weaving ramp scenario is presented in Fig.~\ref{fig:weaving_ramp}. In the simulation, entering vehicles from Lane 0 attempt to find suitable gaps to merge into Lane 1 before reaching the off-ramp, while exiting vehicles from Lane 2 must change lanes twice to reach the off-ramp in time. Meanwhile, through vehicles on Lane 1 must decide, upon entering the interaction zone, whether to change to Lane 2 (bypassing) or remain in their current lane (steadfast).

The main interaction zone, i.e., the segment between the on-ramp and off-ramp, is established based on two key considerations: First, the distance between the on-ramp and off-ramp should not be too long, as this would diminish the intensity of vehicle interactions, reducing the scenario’s representativeness as a weaving ramp. Conversely, a short distance may lead to over-saturated traffic conditions with queue formation, which our model is not designed for.

\begin{figure*}[h!]
\centering
\begin{subfigure}{0.45\textwidth}
    \includegraphics[width=\textwidth]{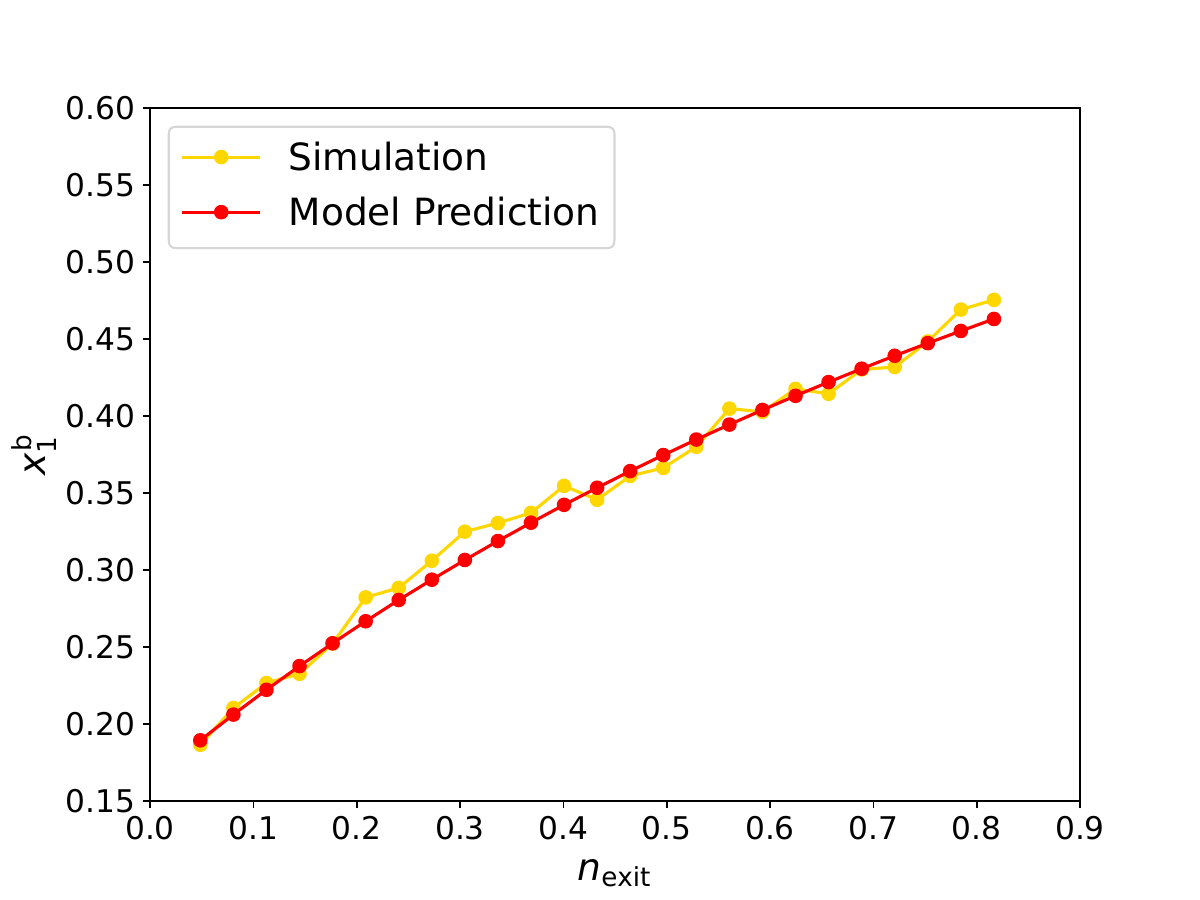}
    \caption{Validation result with $n_\text{enter}=0.1667$}
    \label{fig:first}
\end{subfigure}
\hfill
\begin{subfigure}{0.45\textwidth}
    \includegraphics[width=\textwidth]{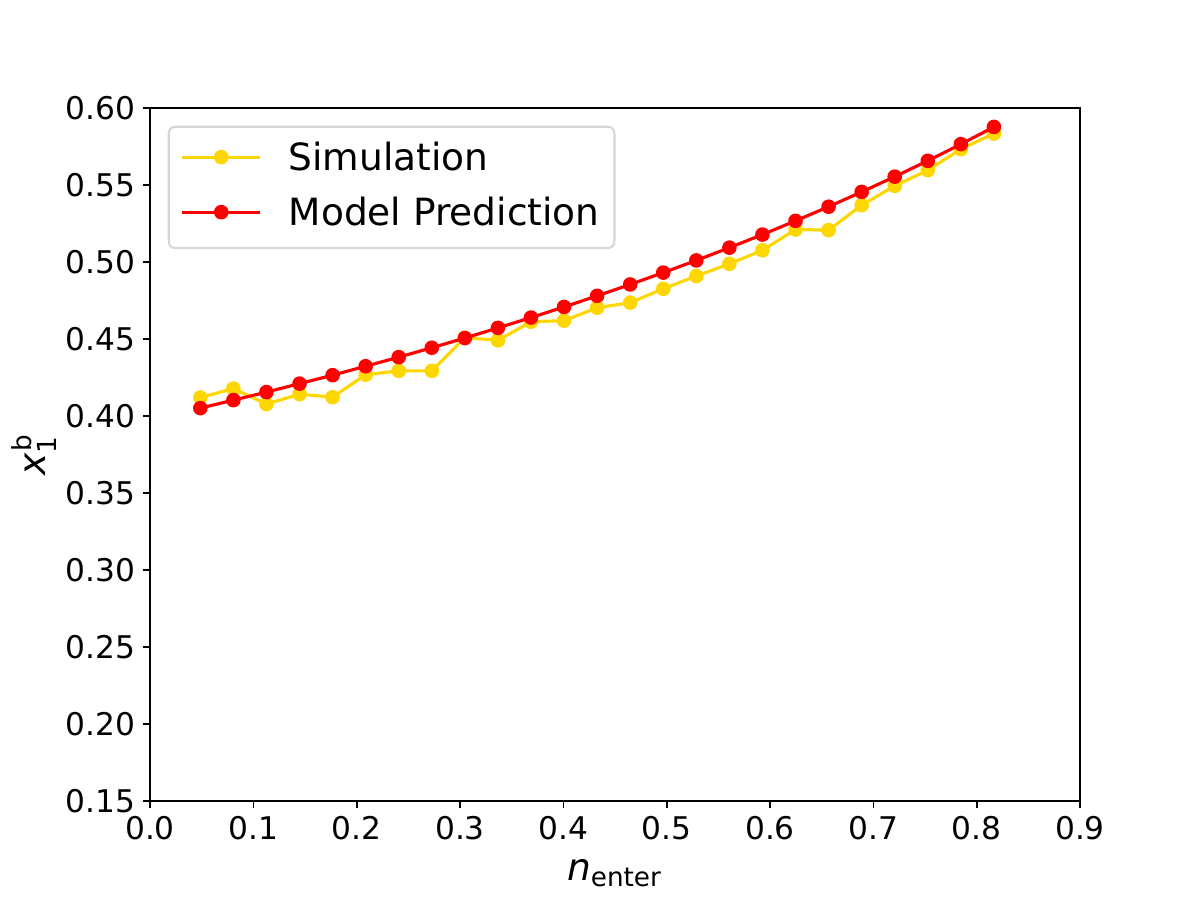}
    \caption{Validation result with $n_\text{2}=0.1667$}
    \label{fig:second}
\end{subfigure}

\vspace{0.5cm} 

\begin{subfigure}{0.45\textwidth}
    \includegraphics[width=\textwidth]{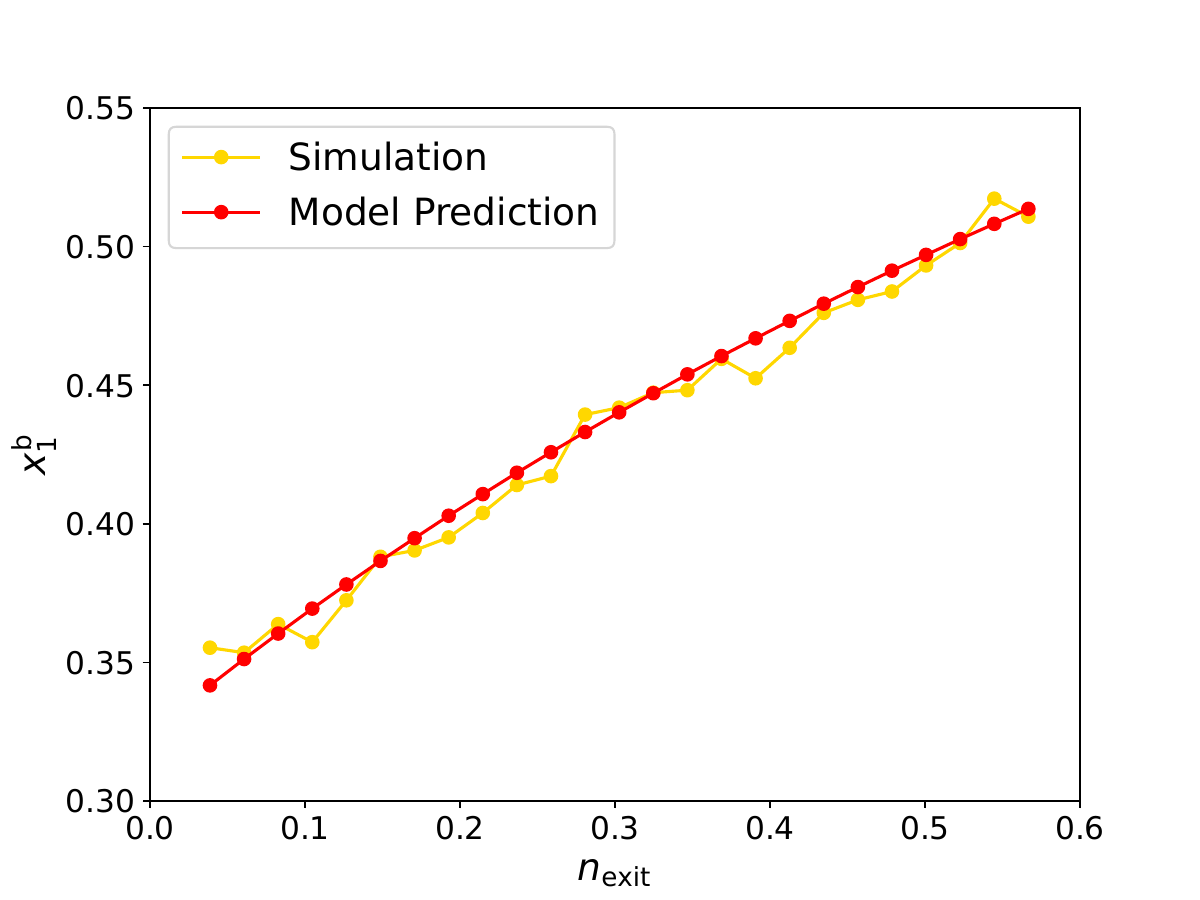}
    \caption{Validation result with $n_\text{enter}=0.4167$}
    \label{fig:third}
\end{subfigure}
\hfill
\begin{subfigure}{0.45\textwidth}
    \includegraphics[width=\textwidth]{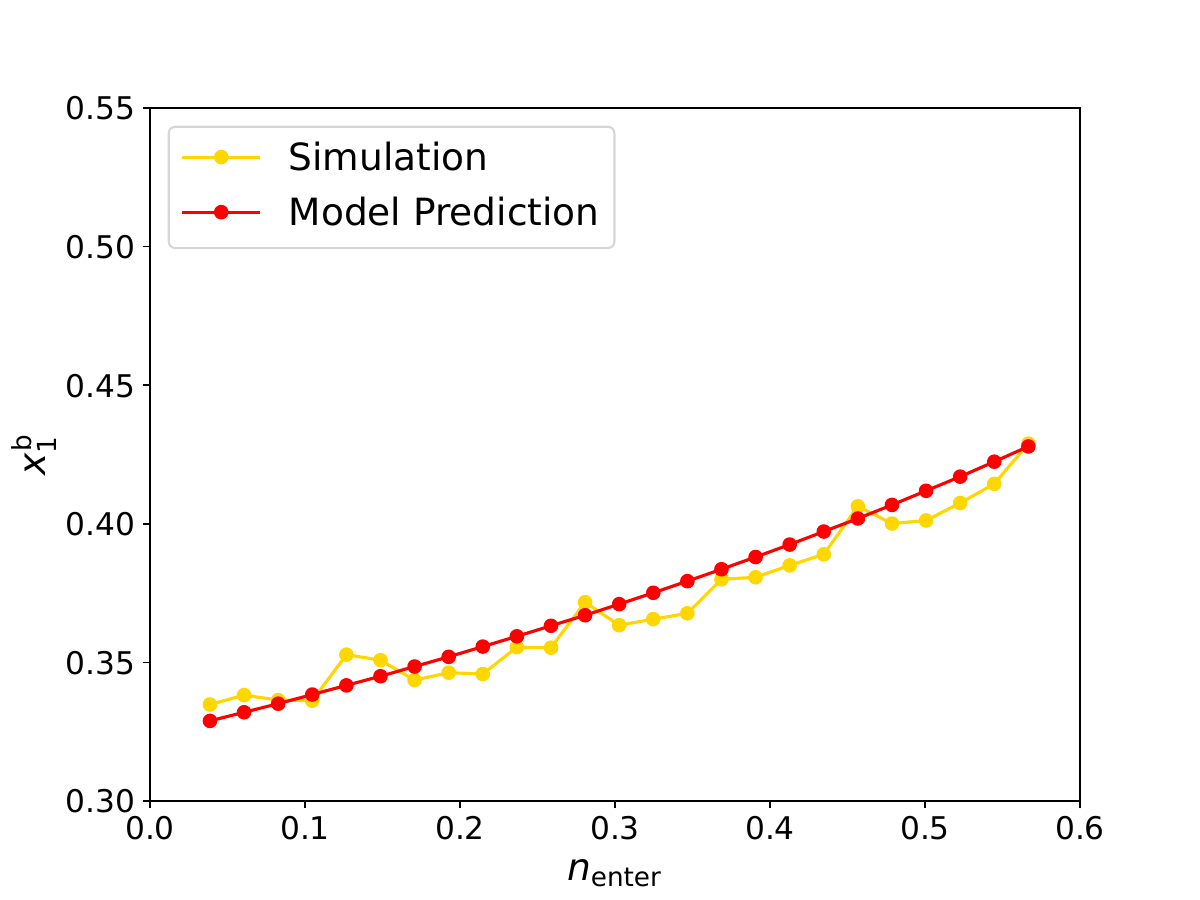}
    \caption{Validation result with $n_\text{2}=0.4167$}
    \label{fig:fourth}
\end{subfigure}

\caption{Validation results demonstrating the strong agreement between our lane choice model and observed simulation outcomes. Despite the cost model's linear structure, the lane choice model successfully captures nonlinear behavior patterns due to the inherent nonlinearity of the equilibrium conditions. Specifically, subfigures (a) and (c) show that an increase in the proportion of exiting vehicles, with a constant flow of entering vehicles, leads to more frequent bypassing behavior by through vehicles on Lane 1, as they shift to Lane 2 to avoid potential congestion. Similarly, a higher proportion of entering vehicles triggers more bypassing behavior due to increased interactions on Lane 1. Subfigures (b) and (d) further reveal that entering vehicles has a more significant impact on bypassing decisions than exiting vehicles. Detailed reasoning can be found in Section~\ref{sec:simulation studies}.}
\label{fig:validation}
\end{figure*}

\subsection{Parameter Calibration}\label{4A}

To evaluate the performance of our model, we first calibrate the cost coefficient vector \(\mathbf{C}\) using the optimization method detailed in~\cite{mehr2021diverge,ruolin2019bifurcating}. A broad range of scenarios is selected for data generation to ensure a comprehensive understanding of the model. Specifically, the simulations are conducted under a total flow rate of 1400 vehicles per hour. In the total flow rate, 600 vehicles per hour is allocated among the three neighboring vehicle types—through vehicles on Lane 2 (\(f_2\)), exiting vehicles (\(f_\text{exit}\)), and entering vehicles (\(f_\text{enter}\)) and the remaining flow rate of 800 vehicles per hour is set for the through vehicles on Lane 1 (\(f_1\)). This ensures sufficient interaction without triggering excessive congestion that could cause a deadlock in the simulation. We further fix the flow rate of one of the three vehicle types—entering vehicles (\(f_\text{enter}\)), exiting vehicles (\(f_\text{exit}\)), and through vehicles on Lane 2 (\(f_2\)) while varying the flow rates of the other two. This results in a total of 415 distinct data points. Each data point captures the equilibrium flow distribution vector \(\mathbf{x} = (x_1^\text{s}, x_1^\text{b})\) for a simulation run with a given flow configuration (\(n_\text{enter}\), \(n_\text{exit}\), \(n_2\)). To ensure that each simulation run reaches a steady state, we set each simulation duration to 20,000 timesteps, with the length of a single timestep as 1 second. All the vehicles have been successfully loaded into the simulation. 

\begin{remark}[Steady state is key to obtaining quality data]
    When calculating the equilibrium flow distribution for each data point, we monitor the ratio of the number of vehicles making different decisions to the total number of through vehicles on Lane 1 accumulating in a time period. The simulation only reaches a steady state if the duration is sufficiently long. Based on our tests, a simulation duration of over 5,000 timesteps is required for the traffic to stabilize. Therefore, we set the duration to 20,000 timesteps, ensuring that the simulation results are as realistic and stable as possible to expose the equilibrium state while maintaining acceptable computational efficiency.
\end{remark}

Solving the calibration optimization problem\cite{mehr2021diverge,ruolin2019bifurcating}, which aims to find the best parameters that enable as many data points as possible to satisfy the condition in Definition \ref{def:wdp_basic}, subject to the constraints on unit costs, we obtain
\begin{align}
    C_1^\text{t} &= C_2^\text{t} = 1, \\
    C_1^\text{m} &= C_2^\text{m} = 1.
\end{align}
We then obtain the calibrated cost coefficients as follows:
\begin{align}
    \alpha &= 1.255, \quad
    \beta = 1.138, \quad
    \omega = 1.000, \\
    \gamma &= 2.384, \quad
    \delta = 3.094, \quad
    \rho = 1.000.
\end{align}

The above parameters reflect the relative impact of different vehicle interactions on the overall cost. Specifically, \(\alpha\) and \(\beta\) are above 1 and $\omega =1$, which suggests that the presence of steadfast vehicles in Lane 1 introduces additional discomfort or delay, possibly due to the increased conflicts and the need to brake and slow down. Further, lane-changing maneuvers of exiting vehicles disrupt the flow in Lane 1, increasing travel time or creating a need for additional adjustments by steadfast vehicles, while entering vehicles merge into Lane 1 without significantly disrupting the traffic flow, possibly due to the adequate space between the ramps leading to a neutral impact on steadfast vehicles. On lane 2, \(\gamma\) is significantly greater than 1, suggesting that the traversing cost for bypassing vehicles in Lane 2 is higher compared to the baseline traversing cost. This could be due to the increased discomfort of needing to maneuver around other vehicles, reflecting the high cost of making a bypass decision. And the higher value of \(\delta\) suggests that conflicting merging movements significantly increase the cost by over three times than the standard merging cost. This reflects the increased difficulty and safety risks associated with such complex maneuvers, making it a significant factor in the decision-making process for through vehicles facing the weaving zone.

\begin{remark}[Parameter values can reflect intrinsic characteristics of the weaving ramp]
    Although the magnitude of the parameter values is determined through calibration, these values can be influenced by various factors such as the geometry of the weaving ramp, speed limits or other intrinsic characteristics of the weaving zone. Thus, the calibrated values should be interpreted in the context of the specific scenario and may vary under different conditions. The effects of changes in speed limit, minimum gap between vehicles, and driver aggressiveness on the calibrated parameters are presented in \ref{4C}.
\end{remark}

\subsection{Model Validation}

While validating our model, we selected a broad range of data to demonstrate its predictive accuracy and robustness. Specifically, we used 320 distinct data points, covering the range of \( n_i \in [0.2, 0.8] \) for \( i \in \{\text{enter, exit, 2}\} \), separate from the data points used for calibration, while ensuring the total demand across these three types of vehicles is fixed at 600 vehicles per hour, with no changes made to other parameters.

Fig.~\ref{fig:validation} presents the validation results, illustrating that the predictions of our lane choice model align closely with the observed simulation outcomes. Subfigures (a) and (c) show that, when the ratio of entering vehicles is held constant, an increase in the proportion of exiting vehicles results in more frequent interactions on Lane 1. This leads to a higher proportion of bypassing behavior by through vehicles on Lane 1, as these vehicles prefer to move to Lane 2 to avoid potential congestion and reduced speeds on Lane 1. Additionally, both subfigures reveal that as the proportion of entering vehicles increases, more through vehicles on Lane 1 tend to bypass. This behavior is driven by similar principles: a higher flow of entering vehicles increases the number of interactions on Lane 1, motivating through vehicles to shift to Lane 2.

Subfigures (b) and (d) provide further insight by holding the flow of through vehicles on Lane 2 constant. When the normalized flow of through vehicles on Lane 2 is fixed, the total proportion of entering and exiting vehicles is constant. Therefore, from the shown increasing bypassing behavior, a scenario with a higher proportion of entering vehicles induces more bypassing behavior by through vehicles on Lane 1 compared to a scenario with more exiting vehicles. In other words, entering vehicles are shown to have a greater impact on the bypassing behavior of through vehicles on Lane 1 compared to exiting vehicles. The phenomenon also corresponds to the high value of $\delta$ in the cost functions as exiting vehicles significantly increase the cost for bypassing.

It is noteworthy that, despite the linear structure of our cost models, the lane choice model can still capture the nonlinear lane choice behavior observed in Fig.~\ref{fig:validation} due to the inherent nonlinearity of the inequalities governing the equilibrium conditions. As a result, our model demonstrates significant flexibility and minimal requirements for computation and calibration while satisfying accuracy, making it a great tool for real-world applications.

\subsection{Model Analysis}\label{4C}

\begin{table*}[!ht]
\caption{Mean Prediction Error Rates (MPER) of the cost function calibrated by the data points in \ref{4A} in different experimental scenarios. The formula for calculating the MPER is: $\text{MPER}=\frac{\sum_{i=1}^{n}\left| \frac{x^\text{s}_{1,\text{S},i}-x^\text{s}_{1,\text{M},i}}{x^{s}_{1,\text{S},i}} \right|}{n}\times 100\%$, in which $n$ is the length of test set; $x^{\text{s}}_{1,\text{S},i}$ is the $i^{\text{th}}$ simulation-generated $x^{\text{s}}_{1}$ in the test set; $x^{\text{s}}_{1,\text{M},i}$ is the $i^{\text{th}}$ $x^{\text{s}}_{1}$ predicted by our model. ${\text{MPER}^{0.1667}_{n_i}}$ represents the experiments with its constant flow \( n_i=0.1667\) for \( i \in \{\text{enter, 2}\} \), while ${\text{MPER}^{0.4167}_{n_i}}$ represents the experiments with its constant flow \( n_i=0.4167\) for \( i \in \{\text{enter, 2}\} \)}
\centering
\begin{tabular}{ccccc}
\hline

Expt.   & ${\text{MPER}^{0.1667}_{n_\text{{enter}}}}$ & ${\text{MPER}^{0.1667}_{n_\text{{2}}}}$   & ${\text{MPER}^{0.4167}_{n_\text{{enter}}}}$& ${\text{MPER}^{0.4167}_{n_\text{{2}}}}$ \\ \hline
\ref{4A}  & 1.15\% & 1.55\% & 1.00\% & 1.05\% \\ 
1-1 & 5.51\% & 3.52\% & 2.41\% & 4.79\% \\ 
1-2 & 5.89\% & 11.00\% & 8.36\% & 1.85\% \\ 
2-1 & 2.54\% & 3.07\% & 3.18\% & 4.30\% \\ 
2-2 & 1.79\% & 2.23\% & 2.74\% & 1.18\% \\ 
3-1 & 1.13\% & 1.26\% & 1.21\% & 0.72\% \\ 
3-2 & 1.15\% & 1.36\% & 1.01\% & 0.87\% \\ 
3-3 & 1.04\% & 1.36\% & 1.10\% & 0.96\% \\ 
\hline
                           
\end{tabular}
\label{table1}
\end{table*}

To test the robustness of our model, we conduct the following three sets of univariate experiments:
\begin{enumerate}
    \item Adjust the velocity limits of through vehicles on lane 2 from 20 m/s to 12.5 m/s (Expt. 1-1) and the velocity limits of exiting vehicles from 20 m/s to 13.9 m/s (Expt. 1-2).
    \item Adjust the minimum gap (the minimum distance a vehicle can maintain from the vehicle ahead) of through vehicles on lane 2 (Expt. 2-1) and entering vehicles (Expt. 2-2) from 2m to 10m.
    \item Increase the vehicle aggressiveness level of through vehicles on lane 2 (Expt. 3-1),  exiting vehicles (Expt. 3-2) and entering vehicles (Expt. 3-3)
\end{enumerate}

We input the data points collected from the experiments into the cost function calibrated by the data points in \ref{4A}. For each test, we set either $n_\text{enter}$ or $n_\text{2}$ as constant, with values of \( n_i \in \left\{0.1667,0.4167\right\} \) for \( i \in \{\text{enter, 2}\} \), following the method used in the model validation section to select data points. From each test, we obtain 100 data points: 50 data points are collected when $n_\text{enter}$ is constant (25 data points for $n_\text{enter}=0.1667$ and 25 data points for $n_\text{enter}=0.4167$), and the remaining 50 data points are collected while $n_\text{2}$ is constant. As the table \ref{table1} shows, all of the mean prediction error rates in different experimental scenarios are no larger than 11\%, which shows that the model is robust.

To explore the practical significance implied by each parameter in the model, thereby enabling informed adjustments under different environmental configurations to maintain accurate model performance, we recalibrate our model using the configuration adjusted in the experiments. The results are shown in Table \ref{table2} and Table \ref{table3}.

\begin{table}[!ht]
\caption{Model parameters under different configurations.}
\centering
\begin{tabular}{ccccccc}
\hline
Expt.    & $\alpha$   & $\beta$  & $\omega$ & $\gamma$  & $\rho$                  & $\delta$        \\ \hline
\ref{4A}            & 1.255    & 1.138   & 1.000           & 2.384     & 1.000            & 3.094    \\ 
1-1              & 1.323    & 2.618   & 1.000           & 2.323 & 1.000            & 6.240  \\ 
1-2 & 1.178 & 2.002 & 1.000 & 2.116 & 1.000 & 8.266\\ 
2-1        & 1.000 & 1.000 & 1.030 & 2.323 & 1.123 & 2.459\\ 
2-2           & 1.236 & 1.000 & 1.000 & 2.204 & 1.044 & 2.726 \\ 
3-1    & 1.302 & 1.056 & 1.000 & 2.436 & 1.000 & 2.958\\ 
3-2         & 1.324 & 1.000 & 1.000 & 2.475 & 1.000 & 2.835  \\ 
3-3    & 1.294 & 1.000 & 1.076 & 2.417 & 1.000 & 2.985\\ 
\hline
                           
\end{tabular}
\label{table2}
\end{table}

\begin{table*}[!ht]
\caption{The fluctuating rate (FR) between model parameters calibrated under different configurations and the configuration we used in \ref{4A} . The formula for calculating the fluctuating rate is: $\text{FR}=\frac{x_{\text{exp}}-x_{\text{ori}}}{x_\text{ori}}\times100\%$. If $\text{FR}>0$, the parameters obtained from the experiments are larger than the corresponding parameters we calibrated in \ref{4A}. Otherwise, if $\text{FR}<0$, the parameters obtained from the experiments are smaller than the corresponding parameters from \ref{4A}. The bold data represent parameters that have significant differences ($\left| \text{FR} \right|>25\%$) compared to the parameters from \ref{4A}. }
\centering
\begin{tabular}{ccccccc}
\hline
Expt.       & $\alpha$    & $\beta$  & $\omega$ & $\gamma$  & $\rho$    & $\delta$  \\ \hline
1-1     & 5.42\%    & \textbf{130.05\%}   & 0.00\%   & -2.56\% & 0.00\%  & \textbf{101.68\%}  \\ 
1-2     & -6.13\% & \textbf{75.92\%} & 0.00\% & -11.24\% & 0.00\% & \textbf{167.16\%}\\ 
2-1     & -20.32\% & -12.13\% & 3.00\% & -2.56\% & 12.30\% & -20.52\%\\ 
2-2     & -1.51\% & -12.13\% & 0.00\% & -7.55\% & 4.40\% & -11.89\% \\ 
3-1     & 3.75\% & -7.21\% & 0.00\% & 2.18\% & 0.00\% & -4.40\% \\ 
3-2     & 5.50\% & -12.13\% & 0.00\% & 3.82\% & 0.00\% & -8.37\%  \\ 
3-3     & 3.11\% & -12.13\% & 7.60\% & 1.38\% & 0.00\% & -3.52\%\\ 
\hline
                           
\end{tabular}
\label{table3}
\end{table*}

The results of experiments show that parameters $\omega$ and $\rho$ have very strong robustness since they gained a 0\% fluctuating rate in most of the experiments, while they only obtained a minimal fluctuating rate in some specific experiments. Parameters $\alpha$ and $\gamma$ also performed robustly, keeping the absolute value of the fluctuating rate below 25\% in all experiments. Parameters $\beta$ and $\delta$ acted robustly in experiments 2 and 3, as all the absolute values of the fluctuating rates are below 25\% in both experiments. Meanwhile, both values gained a high fluctuating rate when the velocity of vehicles in the scenario changed. Given that the results show the changing trends of different parameters under various factors, the model parameters can be adjusted accordingly in new experimental environments to achieve more accurate prediction performance.

In conclusion, we validated the accuracy of our model and further assessed its robustness across different experimental scenarios. By recalibrating the model under various experimental conditions, we identified the variation patterns of model parameters across scenarios with different configurations. These findings enable the model to achieve accurate prediction performance by adjusting the corresponding parameters when applied to unprecedented weaving ramp scenarios not included in our training data. Therefore, our model demonstrates strong practicality when applied to previously unencountered scenarios.



\section{conclusion and future work} \label{sec:future}
In this study, we analyzed the traffic flow patterns in a typical highway weaving ramp scenario, focusing specifically on modeling the lane choice behavior of through vehicles in Lane 1. We developed a game-theoretic framework to capture their cost-minimizing decision-making process. Using the SUMO simulator, we constructed the weaving ramp scenario and conducted simulation experiments to generate data for model calibration and validation. The results demonstrate that our proposed lane choice model accurately captures the behavior of through vehicles in the weaving zone. Additionally, the model offers significant flexibility, minimal computational and calibration requirements, and satisfactory accuracy, making it a practical tool for real-world applications.

For future work, we aim to refine the model by addressing its current limitations and extending its applicability. Specifically, we plan to generalize the model to accommodate more complex geometries, such as multi-lane weaving ramps with intricate interactions, and other types of weaving ramps. A key priority will be to validate and enhance our model using real-world traffic data. Furthermore, we will consider incorporating congestion metrics to capture the impact of traffic density and queues more effectively.

\bibliographystyle{IEEEtran}
\bibliography{main}

\end{document}